# Multimode fiber speckle Stokes polarimeter

YUXUAN XIONG,[1] TING JIANG,[1] HAO WU,[1, *] ZHENG GAO,[1] SHAOJUN ZHOU,[2] ZHAO GE,[1] AND MING TANG[1, *]

[1] *School of Optical and Electronic Information and Wuhan National Laboratory for Optoelectronics, Optics Valley Laboratory, Huazhong University of Science and Technology, Wuhan 430074, China*
[2] *School of Mechanical Science and Engineering, Huazhong University of Science and Technology, Wuhan 430074, China*
*\*Corresponding authors: wuhaoboom@hust.edu.cm; tangming@mail.hust.edu.cn*

**Abstract:** The detection of the state of polarization (SOP) of light is essential for many optical applications. However, it is a challenge for cost-effective SOP measurement due to the complexity of conventional methods and poor transferability of new methods. Here, we propose a straightforward, low-cost and portable SOP measurement system based on the multimode fiber speckle. Convolutional neural network is utilized to establish the mapping relationship between speckle and Stokes parameters. The lowest root mean square error of the estimated SOP on Poincare sphere can be 0.0042. This method is distinguished by its low cost, clear structure and applicability to different wavelengths with high precision. The proposed method is of great value in polarization-related applications.

## 1. Introduction

The estimation of the state of polarization (SOP) is critical in various fields, such as optical communications [1], radar imaging [2], and solar astronomy [3]. Polarimeters, instruments designed to measure the SOP of light, have received significant attention. In the early stage, the focus was primarily on the linearly polarized light, concentrating on measuring the initial three components of the Stokes vector [4,5]. During this period, information about the circularly polarized state, associated with the fourth component, was not obtainable [7,8]. Subsequently, comprehensive polarimeters capable of measuring the full Stokes vector have been investigated. However, conventional SOP measuring methods, such as the rotating quarter-wave plate technique, necessitate the employment of optical elements like polarizers. The inclusion of diverse optical components increases the complexity of the system, and the measurement results are sensitive to environmental noise. Therefore, it is necessary to explore a simple and effective SOP measurement.

Apart from traditional methods, metasurfaces, engineered with designed micro-nanostructures, have been utilized to conduct SOP measurements with enhanced system compactness [9-11]. Nonetheless, they present challenges in design and manufacturing, coupled with limited adjustability and reconfigurability. Furthermore, fibers have been employed to detect SOP, due to their inherent polarization sensitivity [12]. Based on the theory of multi-core polarization interference [13], researchers have utilized three-core photonic crystal fibers for SOP measurement. However, this approach can only detect left- and right-handed polarized light with varying elliptical polarizations. Moreover, the utilization of photonic crystal fibers results in considerable expense. Therefore, there is an urgent need to develop a compact and cost-effective system for measuring the Stokes parameters of light.

Considering the cost and compactness, the multimode fiber (MMF) emerges as an economically practical and lightweight alternative. Not only has it been employed for the sensors of external factors [14], such as temperature and strain, but it has also proven to be an effective medium for measuring the properties of incident light. The large core diameter of MMFs enables the excitation of multiple modes with different phase velocities, and these applications are based on the principle of mode coupling. Coupling of the different modes leads to the speckle patterns with light and dark features at the output. The speckle patterns contain both amplitude distribution and phase information for all modes [15-17]. Changes in the external environment or input light induce the relative phase between modes and affect the speckle [18]. As a result, measurement of input light characterization can be achieved by monitoring and analyzing the speckle patterns generated by the MMF [19].

In this paper, we demonstrated a MMF speckle Stokes polarimeter. Multiple MMF speckles were obtained by changing input SOPs. We employed a convolutional neural network (CNN) to efficiently establish a correlation between SOP and speckle intensity distribution. The system does not require optical alignment and laborious microstructure design. The MMF mitigates system complexity and cost, and enables applicability for different wavelengths. The CNN excludes environmental impacts, thereby enhancing efficiency and flexibility. For applications related to polarization, the proposed method has significant value.

## 2. Principle

The distribution of light intensity emitted from the MMF can be explained by Equation 1, which is a result of different modes coupling of [20]. In the equation, the amplitude, complex phase and effective refractive index of the $i$-th mode are represented by $a_i$, $\exp(j2\pi n_{eff,i}L/\lambda)$ and $n_{eff,i}$. $N$ denotes the number of modes and $L$ is the length of MMF. $\hat{e}_i(x,y)$ represents unit electric field. Any change in the input SOP will alter $a_i$ and $n_{eff,i}$, thereby affecting the intensity distribution [16].

$$I(x,y,L) = \left| \sum_{i=1}^{N} a_i \hat{e}_i(x,y) e^{j\frac{2\pi}{\lambda} n_{eff,i} L} \right|^2 \tag{1}$$

After obtaining speckles related to SOPs from MMF [19], a CNN is employed to establish a mapping relationship between them. The reason for using a CNN is its wide range of applications in image classification and segmentation [21,22]. Research has demonstrated the effectiveness of CNN in complex non-linear modeling and classification tasks for detecting intensity within MMF transmission systems [23]. Furthermore, this approach eliminates the requirement for additional reference structures and solves the problems associated with minor temperature drift in MMF transmission [15]. Therefore, the subsequent task is to denote the SOPs as CNN labels.

$$S_0^2 = S_1^2 + S_2^2 + S_3^2 \tag{2}$$

$$S' = M \cdot S = \begin{bmatrix} S_0' \\ S_1' \\ S_2' \\ S_3' \end{bmatrix} = \begin{bmatrix} m_{00} & m_{01} & m_{02} & m_{03} \\ m_{10} & m_{11} & m_{12} & m_{13} \\ m_{20} & m_{21} & m_{22} & m_{23} \\ m_{30} & m_{31} & m_{32} & m_{33} \end{bmatrix} \cdot \begin{bmatrix} S_0 \\ S_1 \\ S_2 \\ S_3 \end{bmatrix} \tag{3}$$

$$S_{out} = M_1 \cdot M_2 \cdot M_3 \cdots S_{in} \tag{4}$$

Stokes introduced the Stokes parameters (or Stokes vectors) [$S_0$, $S_1$, $S_2$, $S_3$] [24], which are four optically measurable quantities commonly used to describe any SOP. This is because most optical instruments have difficulty in the measurement of high-frequency complex-valued electric and magnetic fields [20]. The four covariates satisfy the relationship shown in Equation 2. For a clear and graphical representation of polarized light, Henri Poincare proposed the Poincare sphere, on which each SOP corresponds to a point. On the basis of Equation 2, an SOP with known Stokes parameters can be considered as a point on a sphere of radius $S_0$ with coordinates ($S_1$, $S_2$, $S_3$), or as a vector that points from the origin (0, 0, 0) to the point ($S_1$, $S_2$, $S_3$), as depicted in Fig. 1(a). However, it is not sufficient to only express the SOP of the light clearly. The effect of the propagation medium on the SOP must also be considered, so that a relational equation between the input Stokes parameters and the output Stokes parameters can be derived. To solve this problem, the Mueller matrix is introduced as a 4×4 order matrix describing the SOP transmission matrix of an optical element [4,5], as shown in Equation 3.

Equation 4 provides the Stokes vector of the output when multiple elements are concatenated in the optical path. This serves as the theoretical value of the output SOP and the label of the CNN.

During the measurement process, there is a natural discrepancy between the predicted and actual vectors, as depicted in Fig. 1(b). The root mean square error (RMSE) is often used as a criterion for the estimation accuracy of this discrepancy [6]. Thus, it can act as the loss function of the CNN. Since the difference in $S_0$ is negligible, the physical meaning of the RMSE is approximated as the length of the difference between the two Stokes vectors, as shown in Fig. 1(b). And its length is utilized to evaluate the accuracy of measurements.

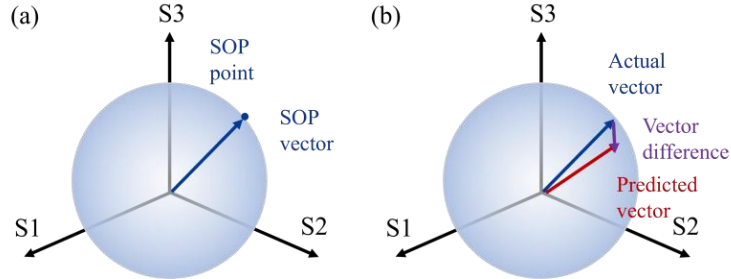

Fig. 1. (a) A SOP with known Stokes parameters can be equated to a point on a sphere of radius $S_0$ with coordinates $(S_1, S_2, S_3)$, or a vector pointing to the point $(S_1, S_2, S_3)$ from the origin. (b) Length of vector difference between actual and predicted SOP vectors are utilized to describe the error of the system.

## 3. Experimental Setup

The experimental setup is shown in Fig. 2. The wavelength of the laser is 1550 nm. The programmable motorized polarization controller (MPC) consists of two autonomous quarter-wave plates as shown in Fig. 3(a). The plates possess the ability to rotate between 0° and 225°, with a precision of 0.225° per step. Later in the text, the degree of rotation will be expressed based on the step of rotation. These dual quarter-wave plates, which can be adjusted independently by software, enable to produce any SOP output to be used as the input of MMF, resulting in a full Stokes parametric measurement. The CCD captures the intensity distribution of the speckle patterns corresponding to different SOPs from MMF, as presented in Fig. 3(b). Once programed, the instruments in this system will automate the data acquisition.

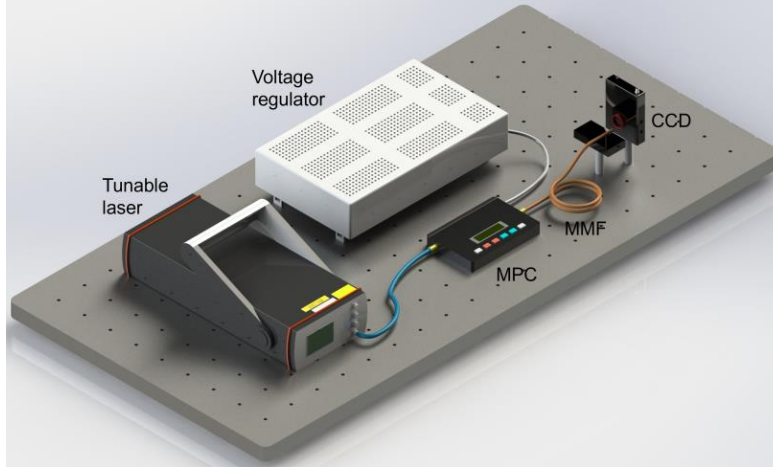

Fig. 2. The scheme of experimental setup for SOP adjustment and speckle collection.

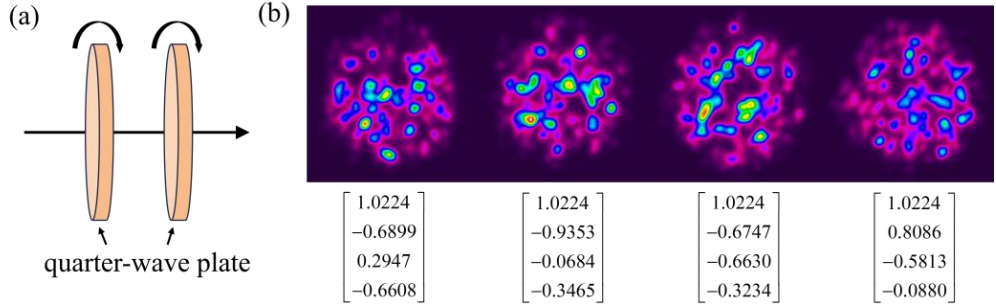

Fig. 3. (a) MPC contains two individually rotatable quarter-wave plates. (b) The speckle patterns of different SOPs are captured by CCD.

The image size of the speckle after cropping is 600×600. To minimize loading time, considerable data has been preprocessed. A 3×3 kernel is employed to perform maximum pooling on the region, resulting in a 200×200 input for the CNN. Through this process, pertinent information is encompassed and loading time for the CNN is reduced. The CNN architecture employed is illustrated in Fig. 4, which consists of several residual blocks, convolution and pooling layers. To better extract detailed information, reduce the number of parameters, and avoid overfitting, the convolutional layers do not change the length or width of the input; it only changes the number of channels, and the dimensionality is reduced mainly through the pooling of the layers. The final fully connected layer generates an estimated value of Stokes parameters.

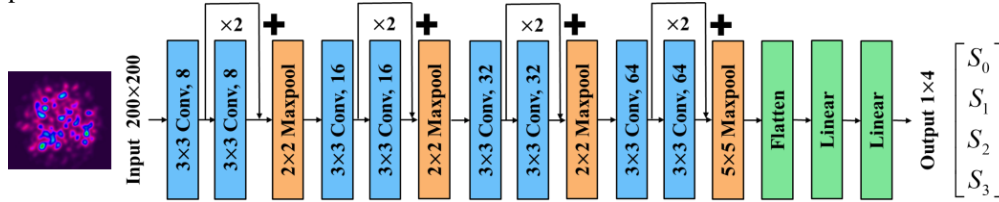

Fig. 4. The structure of CNN mapping SOP to speckle.

## 4. Results and discussion

### 4.1 Full Stokes Parametric Measurement

Investigating the capability of the system required a SOP dataset covering the full Poincare sphere. We regulated the SOP through MPC adjustment with a rotational angle of 4.5° (20 steps), resulting in a SOP distribution traversing the entire Poincare sphere, which is presented in Fig. 5(a). It should be noted that although the rotation of the MPC increases linearly, its impact on the Mueller matrix is non-linear. Therefore, some SOPs in the dataset were sparse while others were dense. Two quarter-wave plates rotated from 0° to 225° individually and 2600 images were collected as a group. To eliminate environmental impact, 16 groups of SOPs that covered the entire Poincare sphere were collected. The dataset was divided into training set, validation set, and test set at an 8:1:1 ratio. In order to evaluate the ability of CNN to predict unknown SOPs, there is no overlap between SOPs in the sets mentioned above. Following the training process, the network generated an RMSE value of 0.0824, which represents the magnitude of the Stokes vector difference between the predicted and actual SOP vectors. For clarity, a distribution function of the loss is presented in Fig. 5(b), while the RMSE of each point is shown in three-dimensional form in Fig. 5(c). Most of the RMSEs are below 0.1, while some exceed 0.2, due to the non-uniform distribution of SOPs. From the results, it can be concluded that, with higher SOP density, the CNN prediction becomes more precise due to the richer information contained within denser SOPs. Conversely, sparse SOPs lead to higher CNN prediction error. Errors of the Stokes parameters are 0.0088, 0.0766, 0.1029 and 0.0721. For clear visualization of the precision of the Stokes parameters prediction, we randomly selected 30 SOPs from the test set and illustrated a comparative graph of their predicted and actual Stokes parameters values ($S_1$, $S_2$ and $S_3$) in Fig. 5(d). The outcome demonstrates the capability of the system to measure the four Stokes parameters.

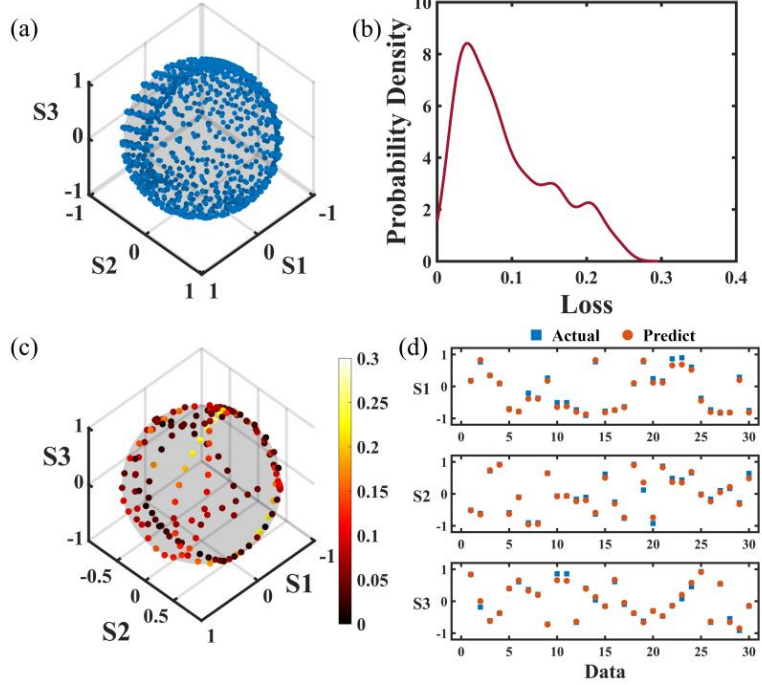

Fig. 5. The dataset of SOPs with a rotation step size of 20 and covering the entire Poincare sphere. (a) Distribution of SOPs on the Poincare sphere. (b) Distribution of RMSE related to SOPs. (c) The errors corresponding to points in test set. (d) Comparison of actual and predicted Stokes parameters.

### 4.2 High Precision Measurement

To improve the estimation accuracy, we reduced the MPC rotation interval to generate denser-distribution datasets of SOPs. The MPC rotation increments are set at 1, 5, 10, and 20 (0.225°, 1.125°, 2.250° and 4.500°). For each interval, 10 sets of 20×20 SOP speckles were collected (Two quarter-wave plates rotate from 0° to 20 times of interval individually as a set.). Fig. 6 illustrates the distribution of SOPs on the Poincare sphere with different rotation steps. The density and homogeneity of the SOPs increase as the interval reduces with same amount of SOPs. Additionally, the speckle changes between adjacent SOPs are not distinguishable to human eyes.

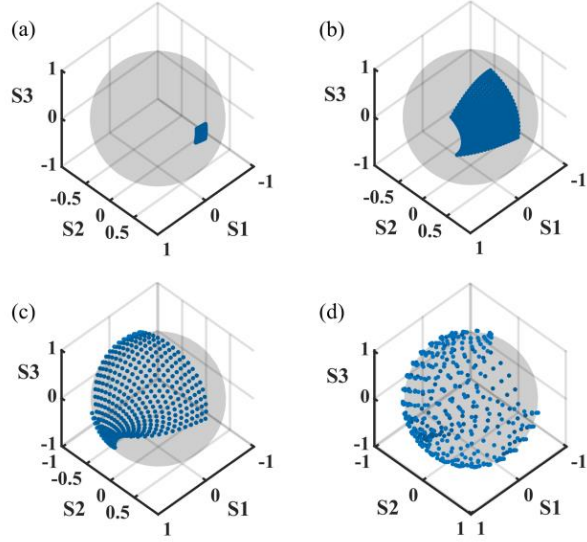

Fig. 6. Distribution of SOP datasets on the Poincare sphere corresponding to different MPC rotation steps. (a) 1 step. (b) 5 steps. (c) 10 steps. (d) 20steps.

Table 1 shows the outcomes for each network. By reducing the MPC rotational angle, the interval between adjacent SOP vectors is diminished resulting in a more uniform and dense SOP distribution, thus enhancing the measurement precision. Fig. 7. depicts the RMSE distribution functions, while Fig. 8. presents the loss distribution in 3D structure for data sets with diverse MPC rotation steps. The concentration of the loss distribution increases and the mean decreases as the MPC rotation angle decreases, indicating continuous accuracy improvement through training data optimization. Furthermore, as the dataset distribution becomes denser and the distribution of neighboring SOPs becomes more linear, the likelihood of sudden large errors decreases, increasing the reliability of measurement. Fig. 9. compares the Stokes parameters of 30 selected SOPs, further illustrating that using data with smaller vector difference intervals increases accuracy. The results obtained in experiments were considerably lower than the stated accuracy. It should be noted that with an MPC rotation step size of 1(0.225°), the variations in vector values among different SOPs are minor, and the RMSE is 0.0042. The errors for the Stokes parameters are 0.0021, 0.0043, 0.0050, and 0.0028, correspondingly. The system measures the four Stokes parameters with similar accuracy and is well-balanced in its sensitivity to each value. This limitation in accuracy is due to the minimum rotation angle of the quarter-wave plate in the MPC. It reveals that the system is capable of achieving highly precise SOP measurement.

**Table 1. RMSE of CNNs with different MPC rotational steps**

| MPC rotational steps | 1 | 5 | 10 | 20 |
|---|---|---|---|---|
| MPC rotation angle interval | 0.225° | 1.125° | 2.250° | 4.500° |
| RMSE | 0.0042 | 0.0082 | 0.0133 | 0.0604 |
| Precision of $S_0$ | 0.0021 | 0.0032 | 0.0091 | 0.0078 |
| Precision of $S_1$ | 0.0043 | 0.0091 | 0.0144 | 0.0861 |
| Precision of $S_2$ | 0.0050 | 0.0086 | 0.0142 | 0.0546 |
| Precision of $S_3$ | 0.0028 | 0.0063 | 0.0068 | 0.0314 |

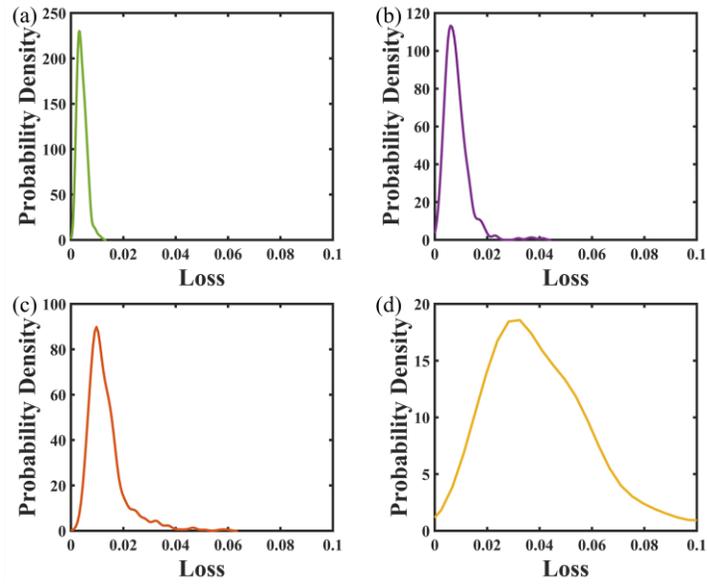

Fig. 7. RMSE distribution of SOPs in test sets with different MPC rotation steps.

(a) 1 step. (b) 5 steps. (c) 10 steps. (d) 20steps.

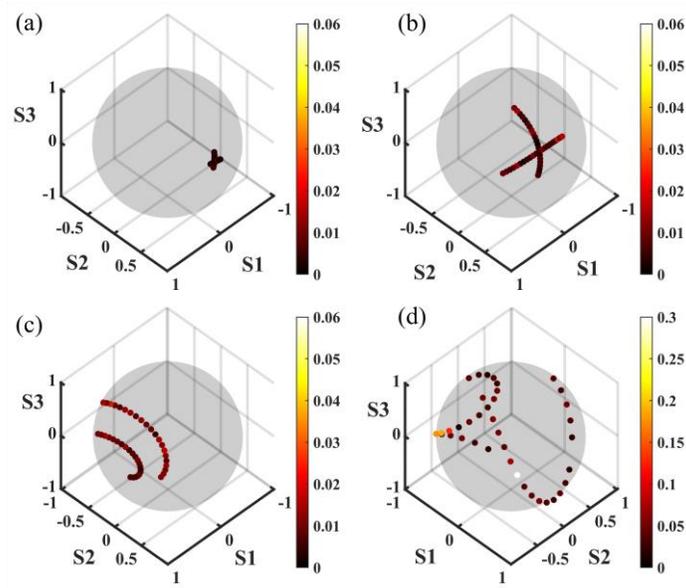

Fig. 8. Loss of SOPs in test sets on the Poincare sphere corresponding to different MPC rotation steps. SOPs were randomly selected and do not overlap with the training or validation sets. (a) 1 step. (b) 5 steps. (c) 10 steps. (d) 20steps.

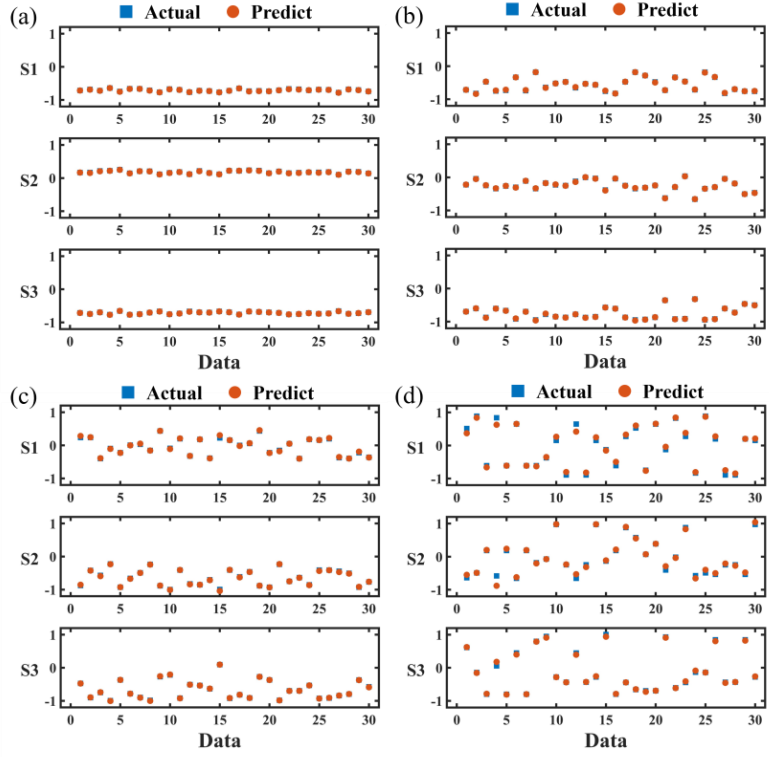

Fig. 9. The comparison of actual and predict Stokes parameters with different MPC rotation steps. (a) 1 step. (b) 5 steps. (c) 10 steps. (d) 20steps.

*4.3 Applicability with Different Wavelengths*

Equation 1 demonstrates that the wavelength influences the intensity distribution of speckles. Therefore, it is imperative to assess the system for its applicability to different wavelengths. The MPC rotation increment was set at 80 (18°), and 10×10 SOP speckles were collected as a set, as shown in Fig. 10(a) (Two quarter-wave plates rotate from 0° to 180° individually as a set.). The laser emits light at single wavelength, ranging from 1545nm to 1555nm in increments of 0.5nm. We collected 5 sets of SOP speckles, covering the entire Poincare sphere for each wavelength. The RMSEs for different wavelengths are depicted in Fig. 10(b). The minimum RMSE is 0.1678 at 1555nm, while the maximum is 0.3400 at 1550nm. Given that the SOPs in the dataset are sparse, this outcome satisfies with the accuracy requirements and the distribution curve of loss appears relatively flat. The results indicate that the largest RMSE occurs at 1550nm, which is the same wavelength used in the previous experiment. Therefore, using other wavelengths in the system can achieve similar SOP measurement accuracy, making it applicable to different wavelength with interest.

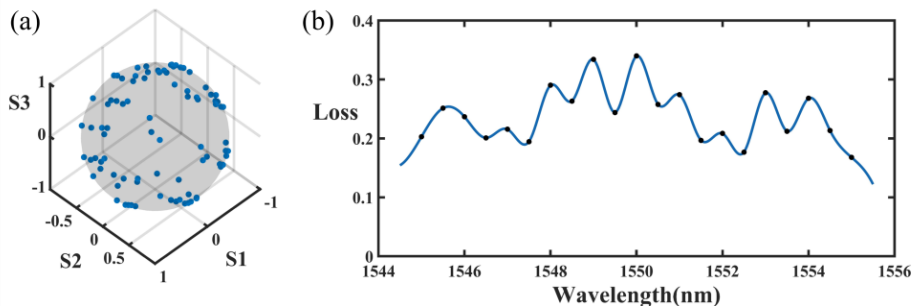

Fig. 10. The datasets include SOPs with a rotation step size of 80, covering the entire Poincare sphere. The wavelength interval is 0.5nm with a bandwidth of 10nm. (a) The distribution of SOPs with different wavelength. (b) Loss of SOPs corresponding to different wavelengths.

## 5. Conclusion

In this paper, we proposed and demonstrated a MMF speckle Stokes polarimeter. The MMF is employed to generate specific speckle patterns with specified SOPs, while a CNN is utilized to establish a mapping relationship between them. We demonstrate the full Stokes parametric measurement capability of the system. And the system can achieve a minimum RMSE of 0.0042. The errors for the Stokes parameters in this case are 0.0021, 0.0043, 0.0050, and 0.0028, respectively. The system achieves highly precise SOP measurement and can be used for different wavelengths. It enables a low-cost and straightforward Stokes polarimeter that provides great value for polarization-related applications.

**Funding.** This work was supported by National Key R&D Program of China under Grant 2021YFB2800902, National Natural Science Foundation of China under Grant 62225110, the Key research and development program of Hubei Province (No. 2022BAA001), and innovation Fund of WNLO.

**Data availability.** Data underlying the results presented in this paper are not publicly available at this time but may be obtained from the authors upon reasonable request.